\documentclass[aps,prl,twocolumn,groupedaddress]{revtex4}

\begin{document}

\title{"Physical quantity" and " Physical reality" in Quantum Mechanics: an epistemological path.}

\author{David Vernette}
\author{Michele Caponigro}

\affiliation{}

\emph{\date{\today}}

\begin{abstract}
We reconsider briefly the \textbf{relation} between
"\textbf{physical quantity}" and "\textbf{physical reality}" in
the light of recent interpretations of Quantum Mechanics. We
argue, that these interpretations are conditioned from the
epistemological relation between these two fundamental concepts.
In detail, the choice as ontic level of the concept affect, the
relative interpretation. We note, for instance, that the
informational view of quantum mechanics ( primacy of the
subjectivity) is due mainly to the evidence of the "random"
physical quantities as ontic element. We will analyze four
positions: Einstein, Rovelli, d'Espagnat and Zeilinger.
\end{abstract}

\maketitle
\section{Introduction}

What do we mean with physical quantities? In quantum mechanics
they play a central role, specifically in a measurement process.
Physical quantities give us information on the state of a physical
system. What do we mean instead with physical reality? We have not
any clear definition. There are many hypothesis on their relation,
most important (Einstein position), was the tentative to establish
a perfect "isomorphism". We are interested to analyze this
possible relation. We retain fundamental this debate, because the
evolution of the two concepts are strictly linked with the
foundations of physical laws. We will utilize the foundations of
Quantum Theory as useful tool to go at the heart of the problem.
\subsection{Realism}
We need to give a "general" definition of "realism". There are
many forms of realism, stronger and weaker. Realism, roughly
speaking, is the belief that there exists an objective world "out
there" independent of our observations. The doctrines of realism
are divided into a number of varieties: ontological, semantical,
epistemological, axiological, methodological. Ontological studies
the nature of reality, especially problems concerning existence,
semantical is interested in the relation between language and
reality. Epistemological investigates the possibility nature and
scope of human knowledge. The question of the aims of enquiry is
one of the subject of axiology, while methodological studies the
best, or most effective means of attaining knowledge. In
synthesis:

\begin{itemize}
\item (ontological):Which entities are real? Is there a mind-independent world?.
\item (semantical):Is truth an objective language-world relation?.
\item (epistemological):Is knowledge about the world possible?.
\item (axiological): Is truth one of the aims of enquiry?
\item (methodological): What are the best methods for pursuing
knowledge.
\end{itemize}

In this paper, we are interested to "ontological realism",
specifically the ontological realism in quantum mechanics. We will
analyze four significative positions: Einstein, Rovelli,
d'Espagnat and Zeilinger. In advance we can say that, starting
from Einstein to Zeilinger, we will assist to a gradual
disappearance of the physical reality (and their relative
isomorphism).

\section{Physical quantity and physical reality in: Einstein, Rovelli, d'Espagnat, Zeilinger}

\subsection{Einstein position\cite{Ei}:} \emph{If, without in any way disturbing a system, we can predict
with certainty (i.e., with probability equal to unity) the value
of a physical quantity, then there exists an element of physical
reality corresponding to this physical quantity}.\\
\\
{\footnotesize This was the basic conjecture of the EPR argument
with the primary objective to prove the incompleteness of QM. The
original paper used entangled pairs of particles states wave,
whose function cannot be written as tensor products. Instead of
using the quite general configuration, usually is considered an
entangled pairs of spin-$\frac{1}{2}$ particles that are prepared,
following Bohm\cite{Bo}, in the so-called \emph{singlet state}
that is rotation invariant and given along any vector by:
\[\Psi(x_1,x_2)=\frac{1}{\sqrt{2}}(| +\rangle _1\otimes|
-\rangle_2-| -\rangle_1\otimes| +\rangle_2)\,\] .}

The above citation lead us to analyze two mentioned fundamental
concepts: (i) physical quantity and (ii) physical reality. We
retain the debate on these two notions completely opened, because
we have not any univocal and deep definition. The importance of
the above statement, to us, is the following strong
epistemological affirmation:\\

\emph{[..]\textbf{then there exists an element of physical reality
corresponding to this physical quantity}}.\\

We note a forced "isomorphism" between two concepts. Through this
line,according us, starts the genuine differences between various
interpretations of quantum theory. Is it correct to "force" the
isomorphic-relation? The relation could be much more complex.\\
We can do some theoretical considerations, first: in a"realist's
world view", there exist physical quantities with "objective
properties", which are independent of any acts of observation or
measurement, but we can not exclude the existence others elements
of physical reality, with a definite values, which do not depend
by measurement. We summarize, below theoretical conjectures:
\begin{itemize}
\item The perfect "isomorphism" between two assumptions ( e.g. Einstein
position)
\item Physical quantity (measurable) without correspondence in the physical
reality (e.g. Zeilinger position) {\footnotesize\item Physical
quantity (measurable) with \emph{"veiled"} correspondence in the
physical reality (e.g. d'Espagnat position)}
\item Unmeasurable physical quantity with possible existence in the
physical reality.
\item Unmeasurable physical quantity with any existence in
the physical reality.
\end{itemize}
Of course, Philosophers can to ascribe these epistemological
positions to philosophical schools. Here, we can easily do many
questions, for instance, (i)what is a physical reality
unmeasurable? (ii)Is it possible that all physical quantities are
measurable? (iii) What is a physical
quantity without the correspondent physical reality? \\
How, we can go out? There are some interesting works, for
instance, the relational quantum mechanics.

\subsection{Rovelli position\cite{Rov}:} \emph{Rovelli departs radically from
such strict Einstein realism, the \textbf{physical reality} is
taken to be formed by the \textbf{individual quantum events}
through which interacting systems (objects)affect one another.
\textbf{Quantum events exist only in interactions} and the reality
of each quantum event is only relative to the system involved in
the interaction. In Relational QM, the preferred observer is
abandoned. Indeed, it is a fundamental assumption of this approach
that nothing distinguishes,a priori, systems and observers: any
physical system provides a potential observer, and physics
concerns what can be said about nature on the basis of the
information that any physical system can, in principle, have.
Different observers can of course exchange information, but we
must not forget that such information exchange is itself a quantum
mechanical interaction. An exchange of information is therefore a
quantum measurement performed by one observing system $A$ upon
another observing system $B$.} \\
These considerations are based on the following basic
concepts\cite{Rov}:\\
\emph{The physical theory is concerned with relations between
physical systems. In particular, it is concerned with the
description that observers give about observed systems. Following
our hypothesis ( i.e. All systems are equivalent: Nothing a priori
distinguishes observer systems from quantum systems. If the
observer O can give a description of the system S, then it is also
legitimate for an observer O' to give a quantum description of the
system formed by the observer O),we reject any fundamental or
metaphysical distinctions as: system / observer, quantum system
/classical system, physical system / consciousness. We assume the
existence of an ensemble of systems, each of which can be
equivalently considered as an observing system or as an observed
system. A system (observing system ) may have information about
another system (observed system). Information is exchanged via
physical interactions. The actual process through which
information is collected and stored is not of particular interest
here, but can be physically described in any specific instance.}
\\

Rovelli position, lead us to think the following epistemological
implications:
\begin{itemize}
\item (i) rejection of individual object
\item (ii) rejection of individual intrinsic properties
\end{itemize}
Some consequence:(a)is not possible to give a definition of the
\textbf{individual} object in a spatio-temporal location, (b)is
not possible to characterize the properties of the objects, in
order to distinguish from the other ones. In other words, if we
adopt the \textbf{interaction} like basic level of the physical
reality, we accept the philosophy of the \textbf{relations} and:
\begin{itemize}
\item (i) we renounce at the possible existence of intrinsic
properties.
\item (ii) and we accept relational properties ( math models).
\end{itemize}
{\footnotesize We remember, for instance, that a mathematical
model based on the relationist principle accept that the position
of an object can only be defined respect to other matter. We do
not venture in the philosophical implications of the relationalism
(i.e. the monism which affirm that there are not distinction a
priori between physical entities). An important advantage of these
approaches is the possibility to eliminate the privileged role of
the observer. This is the importance of Rovelli's approach to
quantum mechanics. In details, Rovelli\cite{Rov} claim that QM
itself drives us to the relational perspective, and the founding
postulate of relational quantum mechanics  is to stipulate that we
shall not talk about \textbf{properties of systems in the
abstract}, but only of properties of systems relative to one
system, we can never juxtapose properties relative to different
systems. Relational QM is not the claim that reality is described
by the collection of all properties relatives to all systems,
rather, reality admits one description per each (observing)
system, each such description is internally consistent. As
Einstein's original motivation with EPR was not to question
locality, but rather to question the completeness of QM, so the
relation interpretation can be interpreted as the discovery of the
incompleteness of the description of reality that any
\textbf{single observer} can give: in this particular sense,
relational QM can be said to show the "incompleteness" of
single-observer Copenhagen QM.}

Rovelli's approach seem do not venture in the clarification of two
notions: physical quantity and physical reality. As we have seen,
he retain fundamental the relation \textbf{between} systems. The
\textbf{math nature} of the relation is the real problem. Of
course, we can ask: math law of what?

\subsection{d'Espagnat position\cite{Es}:} \emph{"defines his philosophical view as "open realism";
existence precedes knowledge; something exists independently of us
even if it cannot be described".} According d'Espagnat, we are
unable to describe the physical reality, but he admit his
\textbf{existence}. For this reason, respect our analysis, is not
clear this position, according d'Espagnat, we can trust only of
physical quantities but we have not any tool to verify their
correspondence in the physical world.

\subsection{Zeilinger position\cite{Ze}:}
The \textbf{individuality} notion have introduced recently radical
interpretation of quantum mechanics. The forced equivalence is
between \textbf{information and individuality} (and not between
physical quantity and physical reality), this is
Zeilinger\cite{Ze}view. He put forward an idea which connects the concept of information with the notion of elementary systems:\\
\emph{First we note that our description of the physical world is
represented by propositions. Any physical object can be described
by a set of true propositions. Second, we have knowledge or
information about an object only through observations. It does not
make any sense to talk about reality without the information about
it. Any complex object which is represented by numerous
propositions can be decomposed into constituent systems which need
fewer propositions to be specified. The process of subdividing
reaches its limit when the individual subsystems only represent a
single proposition, and such a system is denoted as an elementary
system. (qubit of modern quantum physics).}\\

In short, \textbf{random physical quantity} is the main
fundamental rule to fix \textbf{any} correspondence with the
physical reality. Opposite Einstein's position.

\section{conclusion}

We have analyzed, how starting from the genuine realism we have
reached a genuine subjectivism. The physical reality step by step
is \textbf{gradually disappearance}. The physical reality is
replaced by the subject. We ascribe this evolution to the unclear
epistemological relation between physical quantity and physical
reality, so, the interpretation of quantum mechanics is not only
due to the analysis of the formalism. Finally, we conclude with a
paradoxical question: Was Einstein a realist? As we have seen, he
was the only \textbf{real "idealist"} because, he did not give up
to research the physical reality.

\section{{\tiny  }}
{\footnotesize------------------\\ $\diamond$ David
Vernette:Quantum Philosophy Theories
www.qpt.org.uk\\david.vernette.org.uk
\\
$\diamond$ Michele Caponigro: University of Camerino, Physics
Department michele.caponigro@unicam.it}

\end{document}